\documentclass[a4paper,11pt]{article}

\usepackage{wrapfig}
\usepackage{booktabs} 
\usepackage{Definitions}
\usepackage{fullpage}
\usepackage{courier}
\usepackage{authblk}
\usepackage{calc}
\usepackage{mathtools}
\usepackage{enumitem}
\usepackage{multirow}
\usepackage{subfig}
\usepackage{optidef}
\usepackage{algorithm}
\usepackage{algorithmic}

\usepackage{graphicx}
\graphicspath{{./FIG/}}

\usepackage{xcolor}
\newcommand{\red}[1]{{\color{red}{#1}}}

\newcommand{\xhdr}[1]{\vspace{1mm}\noindent{{\bf #1.}}}

\definecolor{C0}{rgb}{0.10588235294117647, 0.6196078431372549, 0.4666666666666667}
\definecolor{C1}{rgb}{0.8509803921568627, 0.37254901960784315, 0.00784313725490196}
\definecolor{C2}{rgb}{0.4588235294117647, 0.4392156862745098, 0.7019607843137254}
\definecolor{C3}{rgb}{0.9058823529411765, 0.1607843137254902, 0.5411764705882353}
\definecolor{C4}{rgb}{0.4, 0.6509803921568628, 0.11764705882352941}
\definecolor{C5}{rgb}{0.9019607843137255, 0.6705882352941176, 0.00784313725490196}

\newcommand{\comment}[1]{\ensuremath{\color{C#1}{C_{#1}}}}

\DeclareMathOperator{\signrankOP}{sign-rank}
\newcommand{\signrank}[1]{\signrankOP(#1)}

\DeclareMathOperator{\signOP}{sign}
\newcommand{\sign}[1]{\signOP(#1)}

\newcommand{\novote}{\ensuremath{\circ}}

\usepackage{pifont}

\usepackage{calc}
\newcommand{\missing}{\makebox[\widthof{$+1$}][c]{?}}

\newcommand{\upvote}{\ensuremath{\uparrow}}
\newcommand{\downvote}{\ensuremath{\downarrow}}

\begin{document}

\title{On the Complexity of Opinions and Online Discussions}
%

\author[1]{Utkarsh Upadhyay}
\author[1]{Abir De}
\author[2]{Aasish Pappu}
\author[1]{Manuel Gomez-Rodriguez}
\affil[1]{%
  {MPI for Software Systems, \{utkarshu, ade, manuelgr\}@mpi-sws.org}
}
\affil[2]{%
  {Yahoo! Research, aasishkp@oath.com}
}


%
%



\date{}

\maketitle

\begin{abstract}
In an increasingly polarized world, demagogues who reduce complexity down to simple arguments based on emotion
are gaining in popularity.
Are opinions and online discussions falling into dema\-goguery?
In this work, we aim to provide computational tools to investigate this question and, by doing so, explore the nature and
complexity of online discussions and their space of opinions, uncovering where each participant lies.

%
%
%

More specifically, we present a modeling framework to construct latent representations of opinions in online discussions
which are consistent with human judgements, as measured by online voting.
If two opinions are close in the resulting latent space of opinions, it is because humans think they are similar.
Our modeling framework is theoretically grounded and establishes a surprising connection between opinions and voting models
and the sign-rank of a matrix.
Moreover, it also provides a set of practical algorithms to both estimate the dimension of the latent space of opinions and
infer where opinions expressed by the participants of an online discussion lie in this space.
Experiments on a large dataset from Yahoo! News, Yahoo! Finance, Yahoo! Sports, and the Newsroom app suggest that unidimensional
opinion models may often be unable to accurately represent online discussions, provide insights into human judgements and opinions, and
show that our framework is able to circum\-vent language nuances such as sarcasm or humor by relying on human judgements instead of
textual analysis.

%
%
%

\end{abstract}

\vspace{-2mm}
\section{Introduction}
\label{sec:introduction}
People join online discussions to, on the one hand, express their own opinions and, on the other hand, approve and
disapprove the opinions expressed by others.
%
%
In this context, there is a wide var\-iety of online platforms that enable their users to approve or disapprove each others'
comments \emph{explicitly} using, \eg, upvotes and downvotes.
%
Here, whenever a user upvotes or downvotes a comment in an online discussion, she reveals the relative position
of her opinion with respect to the opinion expressed in the comment in a latent space of opinions.
By leveraging this observation from multiple comments, upvotes and downvotes, our goal is to investigate the nature and
complexity of an online discussion and its space of opinions, uncovering where each participant lies.

There is a long history of theoretical models~\cite{axelrod1997dissemination, hegselmann2002opinion, holme2006nonequilibrium, yildiz2013binary, yildiz2010voting}
and empirical studies~\cite{choi2010identifying, conover2011political, de2016learning, garimella2018quantifying, guerra2013measure, mejova2014controversy, pang2008opinion}
of opinions.
However, most of this previous work has reduced (potentially) complex opinions down to real-valued numbers---they have assumed that opinions lie
on the real line.
While a unidimensional space of opinions may be sufficient to coarsely characterize people into, \eg, left leaning vs right leaning or liberals vs conservatives,
they may be lacking at accurately representing complex, multisided opinions in an online discussion.
In fact, if such opinions are multisided, a unidimensional representation of opinions will be unable to explain many common voting patterns under some of the most popular
voting models, as we will show in Section~\ref{sec:model}.
That being said, there are two notable exceptions---aspected oriented sentiment analysis~\cite{liu2012sentiment,bross2013aspect} and collaborative filtering based on matrix
factorization~\cite{bennett2007netflix, koren2015advances, linden2003amazon}.
The former relies heavily on ad hoc methods and textual analysis for determining the sides in online reviews rather than on human judgments as we do.
The latter aims to predict users'{} tastes (or \emph{ratings}) about a set of items (\eg, movies) based on a partial observation of user-items ratings. While it typically considers
multiple sides (or factors), it differs from our work in se\-ve\-ral key aspects.
First, users and items typically lie in two different latent spaces while, in our work, comments and voters lie in the same latent space.
Second, it determines the number of dimensions (sides) of the latent spaces empirically (\eg, using cross validation). This is in contrast with our work, which
determines the dimension from first principles.
Finally, most approaches consider real-valued ratings while we consider (binary) votes.

\begin{figure*}[t]
  \centering
  \includegraphics[width=\textwidth]{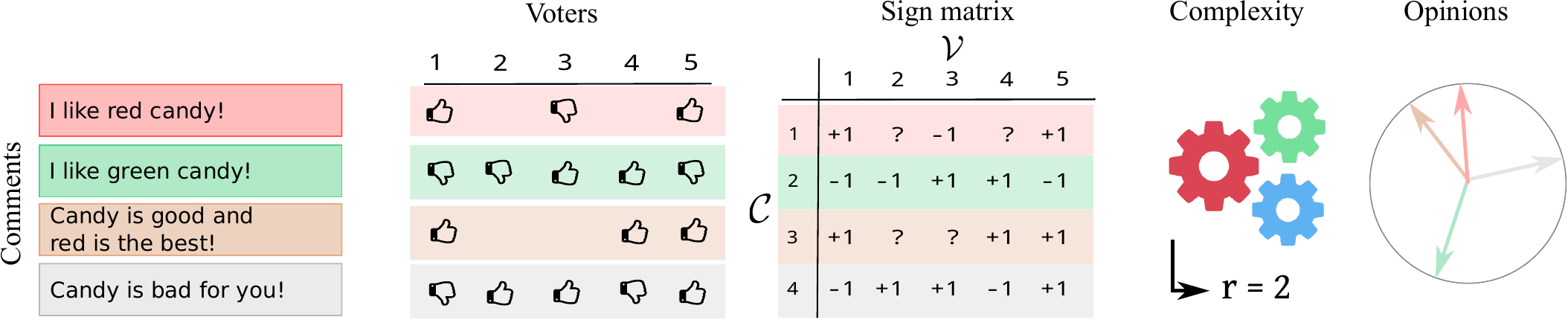}
  \caption{Our modeling framework.
  From left to right, given a (toy) online discussion with a set of comments $\Ccal$ and voters $\Vcal$, our framework
  maps the upvotes and downvotes into a partially observed sign matrix $\Sbb$.
  Within $\Sbb$, each row corresponds to a comment and each column corresponds to a voter. Each $+1$ entry indicates
  that the voter upvoted the comment, $-1$ indicates that she downvoted the comment, and $?$ indicates that the voter did not vote.
  Then, the framework represents the opinions expressed in the comments and those held by the voters as $r$-dimensional real-valued vectors lying in the
  same latent space of opinions.
  Finally, it provides a set of practical algorithms to both estimate the dimension $r$ of the latent space of opinions as well as infer the vectors of
opinions which are consistent with the partially observed sign matrix $\Sbb$.}\label{fig:example-partial-sign-matrix}
\end{figure*}
\xhdr{Current work}
Given an online discussion consisting of a set of comments, which are upvoted and downvoted by a set of voters, we first introduce a latent
multidimensional representation of the opinions expressed in the comments and the opinions held by the voters.
Then, we propose two voting models, one deterministic and another probabilistic, which leverage the above multidimensional representation to
characterize the voting patterns within an online discussion.
Under this characterization, it becomes apparent that the dimension of the latent space of opinions is a measure of complexity of the online
discussion---along how many different axis can the opinions expressed in the comments and the opinions held by the voters differ.
The representation of opinions in this latent space of opinions has a remarkable property: if two opinions are close (far away) in the latent
space it is because the voters---the crowd---think that they are \emph{similar} (\emph{dissimilar}).
Such a property may not hold for other representations of the opinions, \eg, those based only on the textual data in the comments~\cite{le2014distributed,palangi2016deep} because of nuances in the use of language.
Motivated by these observations, we develop\footnote{Implementation at \url{https://github.com/Networks-Learning/discussion-complexity}.}:
\begin{itemize}[noitemsep,nolistsep, leftmargin=0.75cm]
\item[(i)] A polynomial time algorithm to determine an upper bound on the minimum dimensionality that a latent space of opinions needs to have so
that 
they are able to explain a particular voting pattern under the deterministic voting model.
\item[(ii)] An inference method based on quantifier elimination to recover the latent opinions
from the observed voting patterns under the deterministic voting model.
\item[(iii)] An inference method based on maximum likelihood estimation to recover the latent opinions
from the observed voting patterns under the probabilistic voting model.
\end{itemize}

Finally, we experiment with a large dataset from Yahoo! News, Yahoo! Finance, Yahoo! Sports, and the Newsroom app, which consists of one day
of online discussions about a wide var\-iety of topics. Our analysis yields several interesting insights.
We find that only $\sim$$25$\% of the online discussions we anal\-yz\-ed can be explained using a unidimensional representation of opinions,
$\sim$$60$\% of them require a two dimensional representation, and the remaining ones require a greater number of dimensions.
This provides empirical evidence that, to provide opinions representations that are coherent with human judgements, it may be often ne\-cessa\-ry
to move beyond one dimension. 
The presence of multisided opinions is an indication that the discussion may not be falling prey to demagoguery. Such finding is also supported by a positive
correlation between the dimension of a discussion and its linguistic diversity.
Moreover, the estimated $r$-dimensional opinions allow us to predict upvotes/downvotes in a discussion more accurately than a state of the art matrix factorization
method~\cite{mazumder2010spectral} and a logistic regression classifier~\cite{kevin2012machine}.
In this context, whenever an online discussion can be represented using one dimensional opinions, the deterministic model achieves higher predictive performance than
the probabilistic model.
However, for discussions with multisided opinions, the probabilistic model, which allows for noisy voting, provides more accurate predictions.
This suggests that, whenever humans face more complex discussions, their judgements become less \emph{predictable}.
Moreover, we find a positive correlation between the complexity of the discussions and the level of agreement among comments.
Lastly, by looking at particular examples of online discussions, we show that our modeling framework, by relying on human
judgments, may be able to circumvent language nuances like sarcasm and humor, which are often difficult to detect using natural
language processing.
The examples will also illustrate how the dimensions uncover the different \emph{sides} of the discussion.

\vspace{-2mm}
\section{Modeling Opinions and Votes}
\label{sec:model}

At the very outset, the underlying mechanism behind voting on online discussions is fairly commonplace and straight-forward.
Eve\-ry time a user expresses an opinion by posting a new comment in an online discussion, other users can upvote (downvote)
the comment to indicate that they agree (disagree) with the expressed opinion. 
%
%
%
%
%
%

In this context, whenever a user upvotes or downvotes a comment, she reveals the relative position of her opinion with res\-pect to the opinion expressed in the
comment.
By leveraging this observation to multiple comments, upvotes and downvotes, our modeling framework will be able to infer the relative positioning of comments
in an online discussion, as judged by the crowd.
Moreover, by doing so, it will also find a meaningful joint latent representation for the opinions expressed in an online discussion as well as the opinions held by the users
who voted.
In the remainder of the section, we formally introduce our modeling framework, starting from the data it is designed for.
%

\xhdr{Online voting data}
We observe an online discussion con\-sis\-ting of a set of comments $\Ccal$ which are upvoted and downvoted by a set of voters
$\Vcal$.
Here, we keep track of \emph{who} voted \emph{what} by means of the variables $y_{ij} = \{\,\upvote\,, \,\downvote\,, \novote \}$, which indicate that voter $j \in \Vcal$
upvoted, downvoted, or did not vote on comment $i \in \Ccal$, respectively.
Then, we define a (partially) observed sign matrix $\Sbb = [s_{ij}]$, where each $(i, j)$-th entry is given by
\begin{equation}
s_{ij} = \begin{cases}
  +1 & \text{ if }\, y_{ij} = \,\,\upvote\\
  -1 & \text{ if }\, y_{ij} = \,\,\downvote\\
  \phantom{+}? & \text{ if }\, y_{ij} = \novote,
\end{cases}\label{eq:sign-matrix}
\end{equation}
%
%
%
the sign $?$ indicates that the voter did not vote and, thus, we cannot know whether she agrees (or disagrees) with the comment.
We denote by $\Omega$ the set of indexes where we have observations, \ie, $\Omega = \{ (i, j)\, |\, s_{ij} \neq \, \, ? \}$.
Figure~\ref{fig:example-partial-sign-matrix} illustrates the above definitions for a given toy example.
%
%

Next, we introduce our multidimensional representation of the opinions expressed in the comments and those held by the voters and then elaborate on
our voting model, which relates these opinions to the observed voting data.
%
%

\xhdr{Opinion representation} 
Unidimensional (scalar) real-valued representations of opinions have been used most commonly in the literature, owing largely to their interpretability, following the example set by the seminal works of DeGroot~\cite{degroot1974reaching} and Rowley~\cite{rowley1984relevance}.
Thus, we could think of using such unidimensional representation of opinions in our work. However, under that choice, we would be unable to explain certain
voting patterns illustrated below, which are common in many online discussions.

Given an online discussion, assume we represent the opinions expressed in each comment $i \in \Ccal$ as $c_i \in \RR$ and the opinion held by voter $j \in \Vcal$
as $v_j \in \RR$.
Now, we elaborate separately on two of the most popular voting models in the literature~\cite{merrill1999unified}: the proximity model and the directional model.
Under the proximity model, the voters use the Euclidean distance as a similarity measure and decide to cast an upvote if $|v_j - c_i| \le \theta$, where $\theta$
is a threshold, and a downvote otherwise.
Now consider the voting pattern 1 in Figure~\ref{tab:impossible-proximity}. It is easy to show that there are no real-valued scalar opinions $v_1, v_2, v_3$ and
$c_1, c_2, c_3$ leading to such a voting pattern: assume that $v_1 \le v_2 \le v_3$ (as the pattern is symmetric, we can always relabel the voters and comments to make this true) and $c_2 < v_2$. Then $|v_2 - c_2| > \theta \implies v_2 > c_2 + \theta$,
and $|v_3 -  c_2| \le \theta \implies v_3 \le c_2 + \theta$. This contradicts the assumption that $v_2 \le v_3$. We arrive at a similar contradiction with the assumption $c_2 > v_2$.

Under the directional model, the voters use the dot product as a similarity measure and decide to cast an upvote if $v_j \cdot c_i \ge 0$ and a downvote
otherwise. Here, consider the voting pattern 2 in Figure~\ref{tab:impossible-directional}. Again, it is easy to show that there are no real-valued non-zero scalar opinions
$v_1, v_2, v_3$ and $c_1, c_2, c_3$ leading to such a voting pattern. The first row requires $v_1 \cdot c_1 \ge 0$ and $v_2 \cdot c_1 < 0$, which implies $\sign{v_1} \neq \sign{v_2}$.
However, the second row requires $v_1 \cdot c_2 \ge 0$ and $v_2 \cdot c_2 \ge 0$, which implies $\sign{v_1} = \sign{v_2}$ and this leads to a contradiction.

Motivated by the above examples, given an online discussion, we represent the opinions expressed in the comments and those held by the voters as $r$-dimensional
real-valued vectors lying in the same latent space. 
%
%
%
More formally, we represent the opinion expressed in each comment $i \in \Ccal$ as $\cbb_i \in \RR^r$ and we stack all these opinions into a matrix $\Cb$, in which
the $i$-th row corresponds to the opinion $\cbb_i^{T}$.
Similarly, we represent the opinions held by each voter $j \in \Vcal$ as $\vb_j \in \RR^{r}$ and stack all these opinions into a matrix $\Vb$, in which the $j$-th row corresponds
to the opinion $\vb_j^{T}$.
Here, one can think of the dimension $r$ as a measure of the complexity of the online discussion---along how many different \emph{axes} can the opinions expressed
in the comments and the opinions held by the voters differ.
Figure~\ref{fig:example-partial-sign-matrix} illustrates the above definitions using a toy example.
\begin{figure}
  \centering
  \subfloat[Voting pattern 1]{\makebox[0.23\textwidth][c]{%
  \label{tab:impossible-proximity}
    \begin{tabular}{cc|rrr}
     		 & & \multicolumn{3}{c}{$\leftarrow \Vcal \rightarrow$}\\
		          & & 1  & 2            & 3 \\\hline
	\parbox[t]{0mm}{\multirow{3}{*}{\rotatebox[origin=c]{90}{$\leftarrow \Ccal \rightarrow $}}}
          & 1     & \downvote & \upvote & \upvote   \\
          & 2     & \upvote & \downvote & \upvote  \\
          & 3     & \upvote & \upvote & \downvote   \\
  \end{tabular}}}%
  \subfloat[Voting pattern 2]{\makebox[0.23\textwidth][c]{%
          \begin{tabular}{cc|rrr}
     		 & & \multicolumn{3}{c}{$\leftarrow \Vcal \rightarrow$}\\
		          & & 1  & 2  & 3            \\\hline
	\parbox[t]{0mm}{\multirow{3}{*}{\rotatebox[origin=c]{90}{$\leftarrow \Ccal \rightarrow $}}}
               & 1     & \upvote & \upvote   & \upvote  \\
               & 2     & \upvote & \downvote & \upvote   \\
               & 3     & \upvote & \upvote & \upvote   \\
         \end{tabular}\label{tab:impossible-directional}}}
  \caption{Examples of unfeasible voting patterns under the proximity and directional voting models with unidimensional (scalar) real-valued
  representation of opinions.}\label{tab:impossible}
\end{figure}

\xhdr{Voting model}
Given a comment $i$ which expresses an opinion $\cbb_i$ and a voter $j$ who holds an opinion $\vb_j$, we introduce two voting models, one deterministic
and another probabilistic, inspired by the directional model of voting discussed above.

\noindent \emph{--- Deterministic voting model:} In this model, we can uniquely determine each vote $y_{ij}$ from the comment'{}s
opinion $\cbb_i$ and voter'{}s opinion $\vb_j$ by means of the following deterministic rule:
%
\begin{equation}\label{eq:hard-voting}
y_{ij} = \begin{cases}
\,\,\upvote & \text{ if } \inner{\cbb_i}{\vb_j} \ge 0\\
\,\,\downvote & \text{ if } \inner{\cbb_i}{\vb_j} < 0.
\end{cases}
\end{equation}
In the above rule, the vote $y_{ij}$ depends on the \emph{angle} between the opinion vectors $\cbb_i$ and $\vb_j$---if the angle
is greater (less) than $90^\circ$, \ie, $\cbb_i$ and $\vb_j$  lie in the same (different) half-plane in the latent space, then
$y_{ij}=\,\,\upvote \ (\downvote)$.
%
%
%
%
%
%
%

Under this voting model and a partially observed sign-matrix $\Sbb$ derived from votes using Eq.~\ref{eq:sign-matrix}, two natural question emerge:
\begin{itemize}[leftmargin=0.75cm]
\item[(i)] What is the minimum dimension $r$ of the latent space needed to recover the observed entries in $\Sbb$ from the above decision rule
without errors?

\item[(ii)] Once we know the minimum dimension $r$, can we infer the opinion vectors $\cbb_i$ and $\vb_j$?
\end{itemize}
We will answer both questions affirmatively in Section~\ref{sec:sign-rank} and~\ref{sec:embeddings}, respectively.
%

%

\noindent \emph{--- Probabilistic voting model:}
In the definition of our deterministic model, we have implicitly assumed that voters do not make any \emph{errors} while
casting their votes. However, this assumption might be rather restrictive in some scenarios.
To overcome this, we also propose a probabilistic voting model in which votes are binary random variables $Y_{ij}$, and,
\begin{equation}\label{eq:soft}
  \PP[Y_{ij} = y_{ij}] = p(y_{ij}) = \frac{1}{1 + \exp(-s_{ij} \inner{\cbb_i}{\vb_j})},
\end{equation}
where $s_{ij} = +1$ if $y_{ij} =\,\,\upvote$ and $s_{ij} = -1$ if $y_{ij} =\,\,\downvote$.
%
%
%
Similarly, as in the case of the deterministic model, we will propose a method to infer the opinion vectors $\cbb_i$ and $\vb_j$ under this model
in Section~\ref{sec:embeddings}.
In doing so, we will make the assumption that all the latent opinions are finite, \ie, $\exists\,\alpha > 0.\,||\Cb||_{\infty} \le \alpha \, \wedge\, ||\Vb||_{\infty} \le \alpha$.

%

\xhdr{Remark} In the above model definitions, we opt for a similarity metric based on dot products 
because the euclidean distance, used in the proximity model, does not \emph{scale} well with increasing dimensionality:
the relative \emph{volume} of the opinion space where a voter will cast an upvote is proportional to $\left(\theta/\alpha \right)^r$ where $\theta$ is the
threshold for the user, $r$ is the dimension of the latent space and $\alpha$ is the upper bound on the opinion values. 
%
%
%
%

\if{0}
\begin{itemize}
\item \red{There is an additional degree of freedom afforded in this version, because the magnitude of the opinion vectors start playing a role in this setting whereas only their directions mattered in the previous setting.}
\item \red{Also, this model makes the embeddings unique up-to a fixed rotation.}
\end{itemize}
We will discuss how to recover the embeddings in Section~\ref{sec:embeddings}.
\fi

\vspace{-1mm}
\section{Complexity of Online Discussions}
\label{sec:sign-rank}
\begin{figure*}[t]
  \centering
  \includegraphics[width=\textwidth]{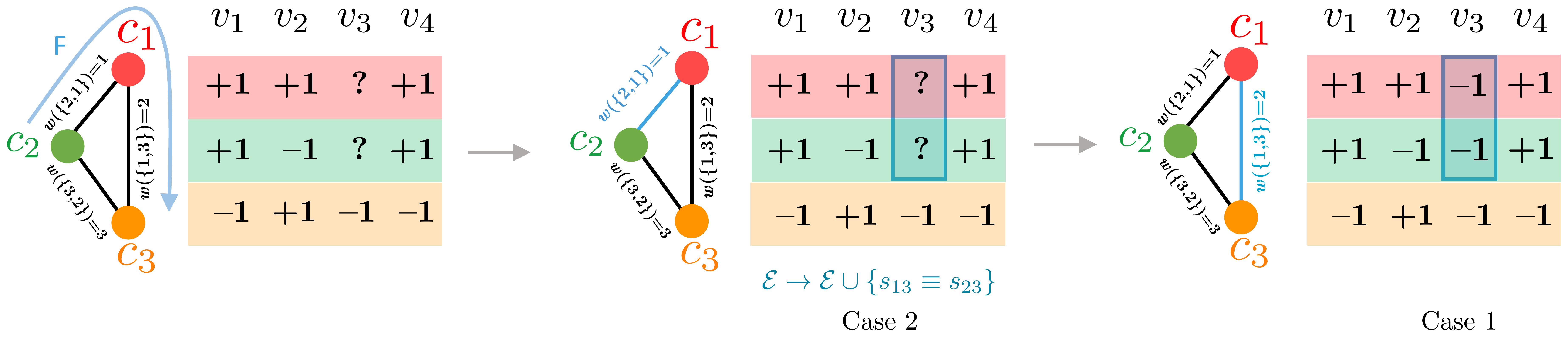}
  \caption{Illustration of Case 1 and Case 2 while selecting the edges of the minimum spanning tree which will help determine the permutation of the rows of the matrix that minimizes $SC^*(\Sbb)$.
    First, the edge $(c_{1}, c_{2})$ is selected and constraint $s_{13} \equiv s_{23}$ is added to $\Ecal$.
  Next, the edge $(c_1, c_3)$ is selected and $s_{13}$ and $s_{23}$ are filled with $s_{33} = -$.
}\label{fig:algo}
\end{figure*}

In this section, we present an algorithm which can determine an upper bound on the minimum dimension $r$ that a latent space
of opinions needs to have so that $\Cb$ and $\Vb$ are able to \emph{explain} voting patterns exhibited by the voters which result in a particular vote-matrix $\Sbb$ under the deterministic
voting model, \ie, $\forall\, (i, j) \in \Omega, \,\, s_{ij} = \sign{m_{ij}}$, where $[m_{ij}] = \Cb \Vb^{T}$.
To this aim, we will first introduce the notion of sign-rank of a sign matrix. Then, we will show that the problem of determining $r$
reduces to finding the sign-rank of a partially observed sign matrix. Finally, we will present an efficient algorithm to estimate the sign-rank.

%



\xhdr{Sign-rank of a sign matrix} Paturi \ea{}~\cite{paturi1984probabilistic} introduced the classical notion of sign-rank of a sign matrix, which is closely related
to the VC dimension of concept classes~\cite{alon2016sign}, as follows:
\begin{definition}
Let $\Mb$ be a real matrix and $\sign{\Mb}$ denote a matrix such that $\forall \, i, j.\, (\sign{\Mb})_{ij} = \sign{\Mb_{ij}}$.
Then, the sign-rank of a sign matrix $\Sbb$ is defined as:
\begin{equation*}
\signrank{\Sbb} = \min \left\{ rank(\Mb)\, |\, \sign{\Mb} = \Sbb \right\}. \label{eqn:sign-rank}
\end{equation*}
%
\end{definition}
Here, we extend the above definition to partially observed sign matrices as follows:
\begin{definition}
The sign rank of a
partially observed sign matrix $\Sbb$ is defined as:
\begin{equation*}
\signrank{\Sbb} = \min \left\{ rank(\Mb)\, | \, \forall\, (i, j) \in \Omega. \, \sign{\Mb}_{ij} = s_{ij} \right\}.
\end{equation*}
%
\end{definition}
It is easy to see that, if the rank of a matrix $\Mb$ is $r$, then we can decompose the matrix into two components
of the form $\Cb \Vb^T$ using, \eg, the singular value decomposition (SVD).
Hence, the problem of determining $r$ reduces to the problem of finding $\signrank{\Sbb}$.


%
Note that the sign-rank of a sign matrix can be much lower than its actual rank, as was noticed by Hsieh \ea{}~\cite{hsieh2012low} in the context of signed graph models.
For example, consider the sign-rank of the matrix $\Bb = 2\II_n - 1_n$, where $\II_n$ is the identity matrix and $1_n$ is the matrix of all $1$ of size $n \times n$.
For $n \ge 3$, $\signrank{\Bb}$ remains $3$ though the matrix itself is always of full rank $n$.
%
%
Moreover, note that, in our setting, the sign-rank does not merely correspond to the \emph{number} of topics being discussed in an online
discussion.
Instead, the complexity may be manifest in the combination of the topics under discussion: the voters may agree with some opinions in a comment while disagreeing
with others.
%


\xhdr{Estimating the sign-rank of a partially observed sign matrix}
The problem of determining whether $\signrank{\Sbb}$ is $1$ can be solved by a simple breath-first search (BFS).
We first create a signed bi-partite graph of comments and voters with adjacency matrix $\Sbb$.
Then for each connected component in the graph, pick one $(i, j) \in \Omega$, set $\cbb_i = +1$ and $\vb_j = s_{ij}$, and fill in the remaining values using BFS by multiplying the source node value with the sign of the edge to arrive at the destination node value.
The intuition is that if voter $j$ has down (up) voted comment $i$, then $i$ and $j$ have opposite (same) polarity.
If a consistent assignment of $\pm 1$ to all the nodes is possible, then $\signrank{\Sbb} = 1$.

However, this algorithm does not generalize to multiple dimensions.
To estimate the sign-rank of a partially observed sign matrix, we adapt the algorithm for (fully observed) sign matrices
proposed recently by Alon \ea{}~\cite{alon2016sign}.
First, we explain the main ideas behind the original algorithm and then describe the necessary, non trivial modifications we propose.

%

The original algorithm upper-bounds the sign-rank of a (fully observed) sign matrix $\Sbb$ by the number of \emph{sign-changes} in the columns
of the matrix.
More formally, define the function $SC(\Sbb)$ as the maximum number of sign changes in any column of the matrix $\Sbb$, \ie,
$SC(\Sbb) = \max_{j} |\{\,i\,|\,s_{i,j}\,\neq\,s_{(i+1),j} \}|$, $Sym(\Sbb)$ as the set of all possible row permutations of $\Sbb$, and
the function $SC^*(\Sbb)$ as:
\begin{equation}
SC^*(\Sbb) = \min_{\Sbb' \in\, Sym(\Sbb)} SC(\Sbb').\nonumber
\label{eqn:sign-change-defn}
\end{equation}
Then, the following lemma establishes the relationship between the sign-rank of matrix $\Sbb$ and $SC^*(\Sbb)$, which the original
algorithm exploits~\cite{alon1985geometrical}:
\begin{lemma}\label{lem:sign-change}
  For a sign matrix $\Sbb$, $\signrank{\Sbb} \le SC^*(\Sbb) + 1$.
\end{lemma}
To use the above result, our algorithm needs to do two tasks:
\begin{itemize}
  \item Find a matrix $\Sbbar = [\sbar_{ij}]$ such that it is a \emph{completion} of $\Sbb$, \ie,
    \begin{equation}
      \sbar_{ij}' = \begin{cases}
        s_{ij} & \text{ if }\, (i, j) \in \Omega\\
        \pm 1 & \text{ if }\, (i, j) \not\in \Omega.\\
      \end{cases}
      \label{eqn:completion}
    \end{equation}
  \item Find $\Sbb' \in Sym(\Sbbar)$ such that it minimizes the maximum number of sign-changes in its columns.
\end{itemize}
The algorithm will output $SC(\Sbb') + 1$ as the estimated sign-rank.

Our algorithm does both tasks together while it computes an estimation of $SC^*(\Sbb)$ using an algorithm by Welzl \ea{}~\cite[See Ex. 4]{welzl1988partition}.
In a nutshell, we construct a graph in which each node corresponds to a row of the partially observed matrix $\Sbb$ and the weight
of each edge between nodes $u$ and $v$ is given by the number of columns where the signs of the corresponding rows disagree.
Then, we extract a spanning tree from the completely connected graph which minimizes the number of sign-changes between pairs
of vertices connected by an edge, as shown in Algorithm~\ref{alg:spanning-tree}.
In the process of creating the spanning tree, we also fill the matrix.
%
%
%
Finally, we derive a permutation of the rows based on the tree to construct $\Sbb'$. 

More in detail, to understand how Algorithm~\ref{alg:spanning-tree} fills the matrix as it computes the spanning tree,
we distinguish two different cases:

\vspace{1mm} \noindent\textbf{Case 1: When, given a column, only one of the rows has a missing entry.}
Consider, for example, we have two rows $u = (+1, +1, \missing, +1)$ and $v = (-1, +1, -1, -1)$.
To calculate the weight of the edge between $u$ and $v$, \ie, $w(\{ u, v \})$ in line~\ref{line:wt} of Algorithm~\ref{alg:spanning-tree}, we
ignore the second column, as $s_{u,2} = s_{v,2}$, and the third column, as it has a missing entry $s_{u,3} = \missing$.
Hence, we report $w(\{ u, v \}) = 2$, because signs of $u$ and $v$ differ in $1$st and $4$th column.
Now, if this edge was chosen in line~\ref{line:argmin} of Algorithm~\ref{alg:spanning-tree}, we \emph{modify} $u$, such that this weight indeed is the
true weight of the edge: we replace $s_{u,3} = \missing$ with the corresponding value in $v$, \ie, $s_{v,3} = -1$, via line~\ref{line:assgn} in Algorithm~\ref{alg:update}.

\vspace{1mm}  \noindent\textbf{Case 2: When, given a column, both entries are unknown.}
If $u = (+1, +1, \missing, +1)$ and $v = (+1, -1, \missing, +1)$, we would still calculate the weight the same way as above. Hence, $w(\{ u, v \}) = 1$.
However, if this edge was chosen in line~\ref{line:argmin} of Algorithm~\ref{alg:spanning-tree}, we could keep the weight the same by merely ensuring that both $u$ and $v$ have the \emph{same} value in the third column, \ie, $s_{u,3} \equiv s_{v,3}$.
Hence, we create and save the constraint that the third column of $u$ and $v$ must always have the same value in line~\ref{line:eqv} of Algorithm~\ref{alg:update}.
Now, say a few steps into the creation of the spanning tree, we find that the missing value in $u$ has to be set to $-1$ as it was being picked as part of an edge under
Case 1 above.
Then, we can also set the same value in the third column of $v$, \ie, $s_{v,3} \leftarrow +1$, via line~\ref{line:assgn} in Algorithm~\ref{alg:update}.
Note that since each column in $\Sbb$ contains at least 1 entry which is $\pm 1$, as each voter has voted at least once, we will eventually hit Case 1 and fill in all missing entries.
This process is illustrated in Figure~\ref{fig:algo}, where the first step creates an equivalence $s_{13} \equiv s_{23}$, and the second step fills in the missing entires with $s_{33}$.
%
%
\begin{algorithm}[t]
  \small
  \begin{algorithmic}[1]
    \REQUIRE A $[|\Ccal|] \times [|\Vcal|]$ sign-matrix $\Sbb = [s_{ij}]$
    \ENSURE {Spanning-tree of the rows.}
    \STATE $l \leftarrow 0$;\quad $F \leftarrow \{\}$;\quad $w: \ZZ^2 \rightarrow \RR$;\quad $Z \leftarrow [|\Ccal|]$;\quad $Y \leftarrow [|\Vcal|]$
    \WHILE{$l < |\Ccal|$}
      \FOR{$u \in Z,\, v \in [|\Ccal|]$; $(u, v) \cup F$ is cycle-free}\label{line:update-loop}
        \STATE $w(\{ u, v \}) \mathrel{+}= | \{ j \in Y\, |\, s_{uj} \neq s_{vj}\,\wedge\,s_{uj}\neq\missing\,\wedge\,s_{vj}\neq\missing \} |$\label{line:wt}
      \ENDFOR
      \STATE $\{ u^*, v^* \} \leftarrow \argmin w(\cdot)$;\quad $\Sbb, Y, Z \leftarrow$ \textsc{Update}$(\{ u^*, v^* \}, \Sbb)$\label{line:argmin}
      \STATE $F \leftarrow F \cup (u^*, v^*)$;\quad $l \leftarrow l + 1$;
    \ENDWHILE
    \RETURN $F$
  \end{algorithmic}
  \caption{Construct a sign-minimizing spanning tree for the columns of $\Sbb$.
  }\label{alg:spanning-tree}
\end{algorithm}
\newcommand{\propagate}{\ensuremath{\,{\scriptscriptstyle \xleftarrow[{\scriptscriptstyle Y,Z}]{\scriptscriptstyle \Ecal}}\,}}
\begin{algorithm}[t]
\small
  \begin{algorithmic}[1]
    \REQUIRE $\{u, v\}$ edge chosen; \AND $\Sbb = [s_{ij}]$ sign-matrix;
    \ENSURE Updated $\Sbb$ and all rows/columns with updated signs.
    \STATE \COMMENT $\Ecal$ initialized as an empty set and persisted across calls.
    \STATE $Z \leftarrow \{\}$;\quad $Y \leftarrow \{\}$
    \FOR{$j \in [|\Vcal|]$}
      \IF{$s_{uj} = \missing$ \AND $s_{vj} \neq \missing$}
        \STATE $s_{uj} \propagate s_{vj}$\label{line:assgn}
      \ELSIF{$s_{uj} \neq \missing$ \AND $s_{vj} = \missing$}
        \STATE $s_{vj} \propagate s_{uj}$
      \ELSIF{$s_{uj} = \missing$ \AND $s_{vj} = \missing$}
        \STATE $\Ecal \leftarrow \Ecal \cup \left( s_{uj} \equiv s_{vj} \right)$\label{line:eqv}
      \ENDIF
    \ENDFOR
    \RETURN $\Sbb,\,Y,\,Z$
  \end{algorithmic}
  \caption{\textsc{Update} procedure used in Algorithm~\ref{alg:spanning-tree}.
  \\ {\scriptsize Operation \propagate{} assigns RHS to all entries equivalent to the LHS in $\Ecal$ (line~\ref{line:eqv}) and updates $Y,\,Z$.}
 }\label{alg:update}
\end{algorithm}

Note that we are \emph{conservative} and greedy while filling in the missing entries, \ie, the edge selected at the $l$th step will have the same \emph{weight} $w(e)$ at the $l$th iteration if the algorithm was to be run on the filled matrix and it will be the minimum weight amongst all valid edges at step $l$ (though the edge may not be unique in having that weight).
Additionally, our algorithm ensures that the weight of the edge selected at iteration $l$ is the minimum possible, given the history of selections.
After obtaining a spanning tree, one can \emph{walk} the tree by performing a depth-first search starting from any source node and create a permutation of the rows by dropping the duplicate nodes in the walk.
Hence, we can obtain $\Sbb' \in Sym(\Sbb)$ and report $SC(\Sbb') + 1$ as $r$.
%

Then, we can establish the upper bound on the dimension by using the following series of self-evident inequalities:
$\signrank{\Sbb} \le \signrank{\Sbb'} \le SC^*(\Sbb') + 1 \le SC(\Sbb') + 1$.

Finally, we would like to highlight that the spanning tree algorithm presented above minimizes the \emph{average} number of sign-changes in
$\Sbb'$. 
Welzl \ea{}~\cite{welzl1988partition} also describe a variant of the algorithm which produces guarantees on the
worst case number of sign-changes in $\Sbb'$; the way the \emph{weight} $w(\cdot)$ is calculated is more involved in the variant.
This variant was used by Alon \ea{}~\cite{alon2016sign} to design the first polynomial time algorithm with approximation guarantees for the sign-rank of the matrix $\Sbb'$.
Remarkably, the \textsc{Update} procedure in Algorithm~\ref{alg:update} can be ported to that variant without any changes, to complete a partially observed matrix $\Sbb$ matrix with worst case guarantees as well.
However, that version is computationally more expensive, more complex, and does not offer significantly better results in practice in our dataset.
Hence, for ease of exposition, we have described the simpler of the two versions.

\xhdr{Computational complexity} The computational complexity of Algorithm~\ref{alg:spanning-tree} can be determined by the computations needed for each missing element in the
matrix $\Sbb$.
Except on the first initialization iteration ($l = 0$), the loop on line~\ref{line:update-loop} will be executed once for each missing element $(u, i)$ in the initial $\Sbb$, right after $s_{ui}$
is fixed via \textsc{Update}.
%
%
%
Hence, the work done for each missing entry, in the worst case, will be incrementing $w(\cdot)$ by 1 $\forall v \in [|\Vcal|]$ such as $s_{uj} \neq s_{vj}$ in the loop on line~\ref{line:update-loop}.
This can be done in $\Ocal(|\Ccal| \log |\Ccal|)$ time if $w(\cdot)$ is implemented as a priority queue.
%
%
As there are at most $|\Ccal| \times |\Vcal|$ missing entries, the computational cost of Algorithm~\ref{alg:spanning-tree} is $\Ocal(|\Ccal|^2 |\Vcal| \log |\Ccal|)$.
All other operations, \ie, initialization of $w(\cdot)$, calculating $\argmin$, checking for cycles in the tree, the creation of an walk, maintaining $\Ecal$, \etc, have lower complexity.

\xhdr{Remark} In our implementation, we do a (non-exhaustive) search over walks with different sources to improve our estimate of $r$ and break ties in calculating the $\argmin{w(\cdot)}$ randomly.
Also, as $\signrank{\Sbb} = \signrank{\Sbb^T}$, we run the algorithm on both matrices and report the smaller value.

\vspace{-2mm}
\section{Multisided Opinion Estimation}
\label{sec:embeddings}
Given an online discussion, we infer the corresponding $r$-dimensional opinions $\Cb$ and $\Vb$ for the deterministic and
probabilistic voting mo\-dels introduced in Section~\ref{sec:model} as follows.


\xhdr{Deterministic voting model}
By definition, under the deterministic voting model, we know that the corresponding $r$-dimensional opinions $\Vb$ and $\Cb$
and the partially observed sign matrix $\Sbb$ need to satisfy the following inequalities:
\begin{equation}
\bigwedge_{(i, j) \in \Omega}
s_{ij} \left( \sum_{k = 1}^r c_{ik}\cdot v_{jk} \right) > 0
\label{eq:existential-sign-rank}
\end{equation}
where $c_{ik}$ and $v_{jk}$ are the $k$-th entry of the opinions $\cbb_i$ and $\vb_j$, respectively, and $\Omega$ is the set of observed
entries in $\Sbb$.
However, we also know from Section~\ref{sec:sign-rank} that, for each voting pattern, there will be a minimum dimension $r_{min}$ under
which such an opinion embedding will not exist.

This reduces the problem of finding the opinions $\Vb$ and $\Cb$ to the existential theory of reals~\cite{tarski1998decision}
%
%
%
%
and, for small values of $r$ and moderate number of comments and voters, $|\Ccal|$ and $|\Vcal|$, this problem can be solved via a procedure
called quantifier elimination~\cite{jovanovic12}, using, \eg, the solver Z3~\cite{de2008z3}.
%
%
%
This procedure eliminates the inequalities in the disjunction in Eq.~\ref{eq:existential-sign-rank} one by one by discovering subsets of $\RR^{r \times (|\Vcal| + |\Ccal|)}$ where they
are satisfied, backtracking to select a different region when an inequality cannot be satisfied, and using sound heuristics to prune the search.
The procedure terminates when \emph{any} assignment to each variable is found or when all regions of $\RR^{r \times (|\Vcal| + |\Ccal|)}$ are eliminated.
%
%
%
%

Note that, if $r < r_{min}$, the solver will conclude that the problem is unsatisfiable. 
Hence, by iteratively increasing $r$ and checking for satisfiability of Eq.~\ref{eq:existential-sign-rank}, one could determine the true sign-rank
of any matrix $\Sbb$.
However, as the most efficient method known for quantifier elimination is doubly exponential in the number of variables, calculating the minimum
dimension $r$ in this way would be computationally more expensive than using the polynomial algorithm introduced in Section~\ref{sec:sign-rank}.
%

\xhdr{Probabilistic voting model} Given a partially observed matrix $\Sbb$, under the probabilistic voting model, we estimate $\Vb$ and $\Cb$ by solving
the following constrained maximum likelihood estimation problem with hyperparameter $\alpha$:
%
\begin{maxi}
{\Cb, \Vb}{ -\sum_{(i,j)\in \Omega}\log \Big(1+\exp\big(-s_{ij}\inner{\cbb_i}{\vb_j}\big)\Big)}
{\label{eq:opt}}{}
\addConstraint{||\Cb||_{\infty}}{\le \alpha}\addConstraint{||\Vb||_{\infty}}{\le \alpha.}
\end{maxi}
%
%
%
%
%
%
%
%
The structure of the above problem allows us to adapt an efficient $1$-bit matrix completion method based on stochastic gradient descend~\cite{bhaskar20151}.
Finally, note that unlike in the deterministic model, for each voting pattern and dimension $r$, there will always exist opinions $\Cb$ and $\Vb$ that best fit the data.

\xhdr{Remark}
In both models, the estimated opinions are unique up-to orthogonal transformations since the inequalities
in Eq.~\ref{eq:existential-sign-rank} and the likelihood in Eq.~\ref{eq:opt} only depend on entries of $\Cb \Vb^T$ and $(\Cb \Ob) (\Vb\Ob)^T = \Cb (\Ob\Ob^T) \Vb^T = \Cb\Vb^T$
for any orthogonal matrix $\Ob$.

\vspace{-1mm}
\section{Experiments}
\label{sec:experiments}

\begin{table}
  \centering
\begin{tabular}{crccc}
\toprule
Dim. &  Discuss. & $|\Ccal|$ & $|\Vcal|$ & Patterns\\
\midrule
$1^{*}$	&	$1{,}139$	&	 $19.7 \pm 16.7$	&	 $19.9 \pm 20.1$	&	 $16.2 \pm 14.8$ \\
$2^{*}$	&	$2{,}820$	&	 $51.5 \pm 85.4$	&	 $57.6 \pm 75.2$	&	 $46.0 \pm 57.3$ \\
$3\phantom{{}^*}$	&	$97$	&	 $43.7 \pm 35.3$	&	 $56.0 \pm 34.5$	&	 $45.4 \pm 24.6$ \\
$4\phantom{{}^*}$	&	$88$	&	 $148 \pm 147$	&	 $195 \pm 194$	&	 $149 \pm 125$ \\
$5\phantom{{}^*}$	&	$245$	&	 $247 \pm 272$	&	 $296 \pm 267$	&	 $218 \pm 179$ \\
$6\phantom{{}^*}$	&	$126$	&	 $445 \pm 461$	&	 $470 \pm 366$	&	 $354 \pm 275$ \\
$7\phantom{{}^*}$	&	$89$	&	 $598 \pm 555$	&	 $706 \pm 518$	&	 $512 \pm 355$ \\
$\ge8$	&	$112$	&	 $2596 \pm 2867$	&	 $2785 \pm 2540$	&	 $1831 \pm 1602$ \\
\bottomrule
\end{tabular}
\caption{%
  Number of comments, voters and unique voting patterns seen in the dataset for discussions with different dimensions.
  The numbers in each column are the mean values $\pm$ the standard deviation.
  {{Dimensions marked with ${}^*$ indicate that they were determined using Z3 and are the true dimensions of the discussions. Our algorithm was used to estimate the dimension of other discussions.}}
%
%
While the dimension of discussions is positively correlated with the size and participants, discussions of different complexity can be found on the entire spectrum, as is also shown in Figure~\ref{fig:overlap}.
}\label{tab:data-description}
\end{table}

\begin{figure}[t]
  \centering
  \subfloat[Sparsity of votes]{\makebox[0.25\textwidth][c]{\includegraphics[width=.25\textwidth]{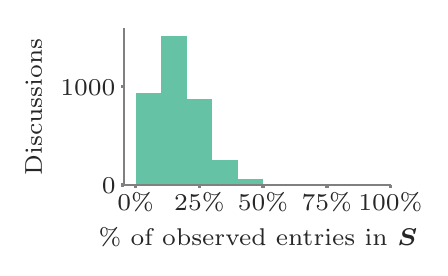}}\label{fig:sparsity-votes}}%
  \subfloat[Unique voting patterns]{\makebox[0.25\textwidth][c]{\includegraphics[width=.25\textwidth]{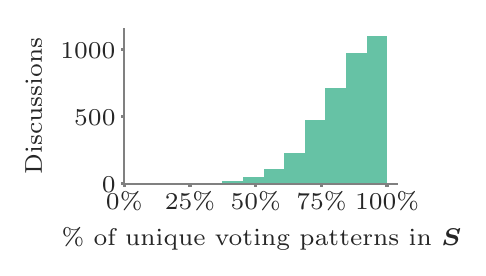}}\label{fig:voting-pats}}%
  \subfloat[Discussions of dim. $2$]{\makebox[0.25\textwidth][c]{\includegraphics[width=0.25\textwidth]{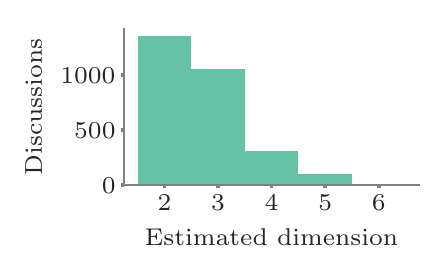}}\label{fig:sign-rank-est-2}}%
  \subfloat[Discussions of dim. $3$]{\makebox[0.25\textwidth][c]{\includegraphics[width=0.25\textwidth]{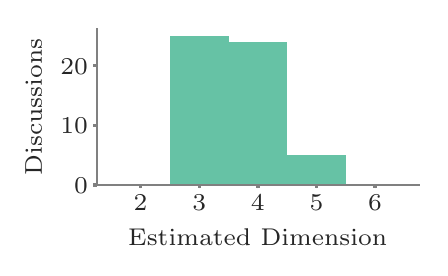}}\label{fig:sign-rank-est-3}}%
  \caption{Distribution of number of observed elements and fraction of unique voting patterns in matrix $\Sbb$ (Panel a and b) and
performance of our algorithm for dimension estimation (Panel c and d).
Panel (a) shows that $\Sbb$ for most discussions is very sparse and Panel (b) shows that many $\Sbb$ have overlapping voting patterns.
Panel (c) and (d) show that for discussions whose opinions can be explained using two (three) dimensions, our algorithm recovers the true dimension for $48\%$ ($46$\%) of the discussions and is off by one for $37\%$ ($44\%$) of them.}
  \label{fig:sign-rank-est-eval}
\end{figure}

\xhdr{Data description}
Our dataset contains $\sim$$19{,}800$ online discussions, each associated to an article from Yahoo! News (including contributed articles), Yahoo! Finance, Yahoo!
Sports, and the Newsroom app, which contain $\sim$$5$ million votes, cast by $\sim$$200{,}000$ voters on $\sim$$685{,}000$ comments, posted by $\sim$$151{,}000$ users.
These votes were randomly sampled from all votes which were cast on comments made by users in the US on August 8, 2017.
%
%
%
%

As a pre-processing step, we discard discussions with less than 10 comments, as they contain too little data to provide meaningful results.
After this step, our dataset consists of $4{,}700$ discussions, with $\sim$$4.5$ million votes, cast by $\sim$$199{,}000$ voters, on $\sim$$645{,}000$ comments,
posted by $\sim$$137{,}000$ users.
Figures~\ref{fig:sparsity-votes} and~\ref{fig:voting-pats} show the \emph{richness} of the data in the votes gathered in the online discussions by means of the sparsity of $\Sbb$ and the
number of unique columns of $\Sbb$, which we name as \emph{voting patterns}.
%

%

%

%
%

%

\begin{figure}[t]
  \captionsetup[subfigure]{singlelinecheck=true,margin=5pt,hangindent=10pt,indent=10pt}
\centering
  \subfloat[$|\Ccal|$ with dim = 2]{\makebox[0.25\textwidth][c]{
    \includegraphics[width=0.17\textwidth]{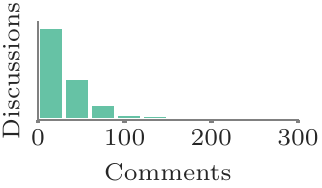}}\label{fig:overlap-1}}%
  \subfloat[$|\Ccal|$ with dim = 4]{\makebox[0.25\textwidth][c]{
    \includegraphics[width=0.17\textwidth]{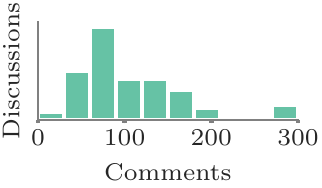}}\label{fig:overlap-2}}%
  \subfloat[$|\Ccal|$ with dim = 6]{\makebox[0.25\textwidth][c]{
    \includegraphics[width=0.17\textwidth]{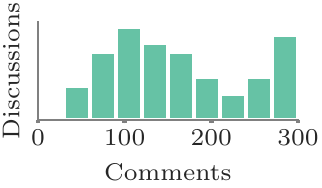}}\label{fig:overlap-3}} 
\subfloat[Lexical similarity]{\makebox[0.25\textwidth][c]{\includegraphics[width=0.23\textwidth]{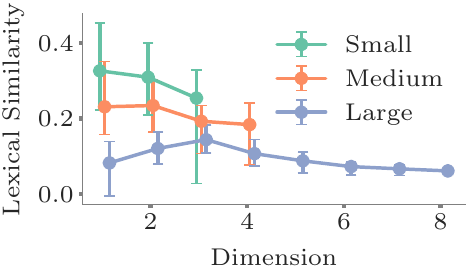}}\label{fig:lexical-disagreement}}
\caption{Panels (a), (b), and (c) show the distribution of the number of comments $|\Ccal|$ per discussion for different dimension values. The distributions of the number of voters and voting patterns show similar spread.
Panel (d) shows the lexical similarity of an online discussion against its dimension. The lexical similarity is the mean Jaccard similarity of the lexical tokens used in all pairs of comments in the discussion.
  The discussions are classified by the number of comments as small (smallest $33\%$), medium, and large (largest $33\%$) and data is shown only for dimensions which contain
  more than 5 discussions.}
  \label{fig:overlap}
\end{figure}

\xhdr{Complexity of discussions}
%
%
In this section, we compute the complexity of the discussions, \ie, the dimensionality of the latent space of opinions, for the online discussions in our dataset.
For each online discussion, we determine whether it can be explained using an unidimensional space of opinions using the linear time algorithm presented at the beginning of Section~\ref{sec:sign-rank}.
If it cannot be explained using one dimension, we determine whether it can be explained using a two- or three-dimensional space of opinions via
quantifier elimination\footnote{In practice, we found quantifier elimination to be sufficiently scalable to test whether an online discussion can be explained
using up to two dimensions.}, following Section~\ref{sec:embeddings}.
Finally, if it cannot be explained using two or three dimension, we resort to the algorithm presented in Section~\ref{sec:sign-rank}, which
provides an upper bound on the true dimension.

Table~\ref{tab:data-description} summarizes the results, which show that the opinions of about $1{,}139$ ($25$\%) of the discussions can be explained
using one dimension, $2{,}820$ ($60$\%) of the discussions require two dimensions, while the remaining $757$ discussions ($15$\%) require a higher
number of dimensions. This allows us to conclude that the opinions in most of the online daily discussions ($85$\%) can be explained using a latent space
of relatively low dimensions, \ie, $r \leq 3$.
Moreover, while discussions with a higher number of participants ($|\Ccal|+|\Vcal|$) and richness (\ie, higher number of voting patterns and lower sparsity)
require, in general, a latent space of opinions with a larger number of dimensions, there is a large variability spanning the entire spectrum of online
discussions, as shown in Figures~\ref{fig:overlap-1},~\ref{fig:overlap-2}, and~\ref{fig:overlap-3}.
%

Next, we evaluate how tight is the upper bound on the true dimension provided by our algorithm for online discussions, which we used above
for discussions whose dimension we could not find using quantifier elimination.
To this aim, we run our algorithm on discussions whose true dimension we could find using quantifier elimination and compare the upper bound
with the true dimension.
%
%
Figure~\ref{fig:sign-rank-est-eval} summarizes the results, which show that, for discussions whose opinions can be explained using two (three) dimensions,
our algorithm recovers the true dimension for $48\%$ ($46$\%) of the discussions and is off by one for $37\%$ ($44\%$) of them.
%
%
%

Finally, we investigate the relationship between the complexity of the discussions, estimated using human judgments, and their linguistic diversity,
estimated using textual features. To this aim, for each online discussion, we compute the average Jaccard similarity of the lexical tokens used in all
pairs of comments as a measure of lexical similarity.
Figure~\ref{fig:lexical-disagreement} summarizes the results, which show a positive correlation between the complexity of a discussion and its linguistic
diversity, as one may have expected.

\begin{figure}[t]
  \captionsetup[subfigure]{singlelinecheck=true,margin=5pt,hangindent=10pt,indent=10pt}
\centering
\subfloat[Discussions of dim.~$1$]{\makebox[0.25\textwidth][c]{\includegraphics[width=0.25\textwidth]{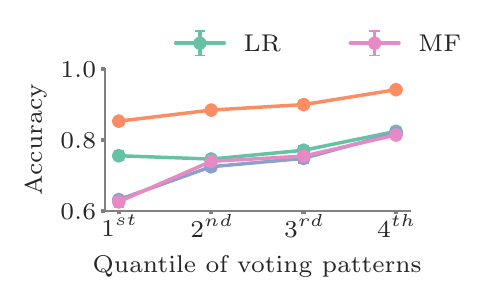}}\label{fig:prec-at-10-1d-sat}}%
\subfloat[Discussions of dim.~$2$]{\makebox[0.25\textwidth][c]{\includegraphics[width=.25\textwidth]{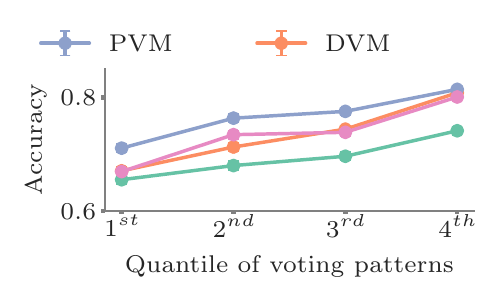}}\label{fig:prec-at-10-1d-unsat}}%
\subfloat[Agreement vs upvotes]{\makebox[0.25\textwidth][c]{\includegraphics[width=0.25\textwidth]{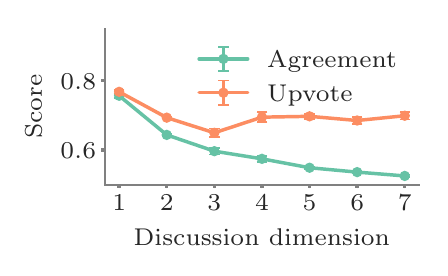}}\label{fig:rel-disagreement}}%
\subfloat[Distribution of Agreement/upvote]{\makebox[0.25\textwidth][c]{\includegraphics[width=0.25\textwidth]{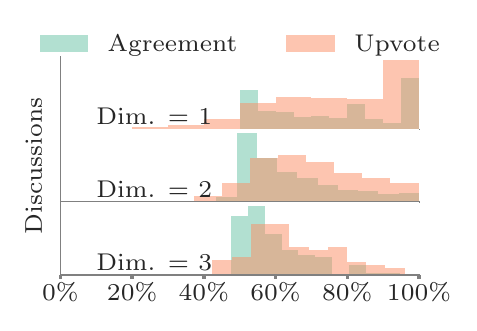}}}
\caption{Panel (a) and (b) show vote prediction accuracy for the deterministic voting model (DVM), the probabilistic voting model (PVM), a state of the art matrix factorization method~\cite{mazumder2010spectral} (MF), and a logistic regression
classifier~\cite{kevin2012machine} (LR) using textual features extracted using Rake~\cite{rose2010automatic}.
The performance for DVM, PVM and MR uniformly increases as the number of unique voting patterns increases, in contrast, the performance for LR remains relatively constant.
%
%
%
Panel (c) and (d) show agreement and percentage of upvotes among all votes in online discussions.
Agreement is measured in terms of percentage of comment pairs $(\cbb_i, \cbb_j)$ for which $\cbb_i^{T} \cbb_j > 0$. The higher the dimension of the latent space of opinions, the lower the agreement between comments, however, such finding would not be apparent directly from the fraction of upvotes, which remains relatively constant.
}\label{fig:test-acc}
\end{figure}

\xhdr{Opinions in online discussions}
In this section, we first evaluate both quantitatively and qualitatively the \emph{quality} of the estimated $r$-dimensional opinions in the online
discussions and then leverage the estimated opinions to shed some light on the level of controversy in online discussions.
Here, we used the opinion estimation method for the probabilistic voting model introduced in Section~\ref{sec:embeddings}, which scales graciously
with the dimension $r$.

In terms of quantitative evaluation, we assess to which extent the deterministic voting model (DVM) and the probabilistic voting model (PVM)
can predict whether a voter will upvote or downvote a comment from the estimated opinions in comparison with two baseline methods: (i)
a state of the art matrix factorization method~\cite{mazumder2010spectral} (MF), which assumes the entries in $\Sbb$ are real valued,
and (ii) a logistic regression classifier~\cite{kevin2012machine} (LR) that uses $200{,}000$ keywords extracted using Rake~\cite{rose2010automatic} as features.
To this aim, for each discussion, we held out some of the observed upvotes and downvotes, estimate the opinions from the remaining votes, and
then predict the votes from the held-out set.
However, since our data is very sparse, as shown in Figure~\ref{fig:sparsity-votes}, and even holding out a small fraction of votes may change the underlying
dimension of the latent space of opinions, we resort to leave-one-out validation.
Moreover, we randomly select $200$ discussions to tune the hyperparameters of the probabilistic voting model and these discussions are excluded
from the validation set.
Figure~\ref{fig:test-acc} summarizes the results, which show that:
\begin{itemize}[leftmargin=0.75cm]
\item[(i)] DVM (PVM) beats all other methods for discussions with dimension 1 (2).
\item[(ii)] The performance of DVM, PVM and MF increases as the number of unique voting patterns in the dataset increase, in contrast,
the performance of LR, which uses text features, does not benefit much from additional voting patterns.
\item[(iii)] While for discussions where opinions can be explained using two dimensions, PVM achieves better performance,
for discussions which require only one dimension, DVM beats PVM.
A potential explanation for this behavior is that, whenever humans face simpler decisions, \ie, their opinions can be explained using one dimension, they become more \emph{predictable}.
\end{itemize}

\newcommand{\framecomment}[2]{\framebox[\textwidth][l]{\makebox[1cm][c]{\comment{#1}:}\parbox{\textwidth-1.5cm}{#2}}}
\newcommand{\omitted}{{\scriptsize\color{gray}[\dots]}}
\begin{figure*}
\captionsetup[subfigure]{labelformat=empty}
\centering
\begin{minipage}[c]{0.20\textwidth}
 \subfloat[Estimated opinions]{\makebox[\textwidth][c]{\includegraphics[width=\textwidth]{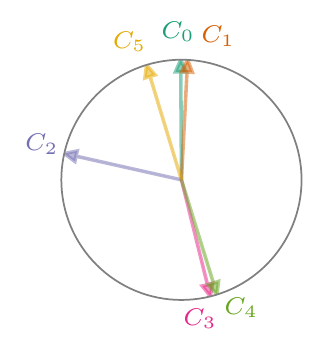}}\label{fig:embedding-example-2}}\\
  \subfloat[Sentiments]{\makebox[\textwidth][c]{\includegraphics[width=\textwidth]{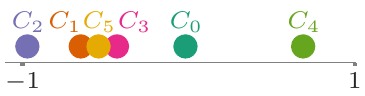}}\label{fig:embedding-example-2-sent}}
\end{minipage}%
\begin{minipage}[c]{0.75\linewidth}
\framecomment{0}{\omitted{} [Donald Trump] has \omitted{} Enquirers\footnote{\emph{National Enquirers} is a well known entertainment magazine in US.} [which] he considers a treasure trove of information.}
\framecomment{1}{He should change his name to Donald J Dubious.}
\framecomment{2}{\omitted{} Trump can be an \#\$\%\$, and Islam can be cancer \omitted{} they are not mutually exclusive \omitted{}}
\framecomment{3}{Why not? Try anything. Terrorism has got to stop now!}
\framecomment{4}{It is a great idea}
\framecomment{5}{Trump family motto-``It's not a lie if you believe it.''}

\end{minipage}
\caption{A subset of comments and estimated opinions for an online discussion about politics.
Two pairs of comments, (\comment{0}, \comment{1}) and (\comment{3}, \comment{4}), express a similar opinion, however, the lexical overlap between comments within each
pair is low. Remarkably, our method is able to identify they are similar, as a human would do, by leveraging the judgements of the voters, and their estimated
opinions lie close to each other in the latent space of opinions.
Moreover, the estimated opinion of a comment expressing an opposite view to the ones above, \comment{2}, lies in an orthogonal direction.}\label{fig:embeddings-2}
\end{figure*}

In terms of qualitative evaluation, we first assess to which extent comments in online discussions agree (or disagree) by analyzing the estimated opinion embeddings of the comments. More
specifically, for each online discussion, we compute the percentage of distinct comment pairs $(\cbb_i, \cbb_j)$ for which $\cbb_i^{T} \cbb_j > 0$ and compare this quantity with the
percentage of upvotes among all votes (upvotes and downvotes).
Figure~\ref{fig:rel-disagreement} summarizes the results, which show that the higher the dimensionality of the latent space of opinions, the lower the agreement
between comments, as one may have expected. Remarkably, such finding would not be apparent directly from the relatively constant fraction of upvotes. However,
relative upvotes are typically the measure of \emph{controversy} (or, rather, consensus) employed by various websites, like Reddit, to sort articles/comments.

\begin{figure*}
\captionsetup[subfigure]{labelformat=empty}
\centering
\begin{minipage}[c]{0.20\textwidth}
 \subfloat[Estimated opinions]{\makebox[\textwidth][c]{\includegraphics[width=\textwidth]{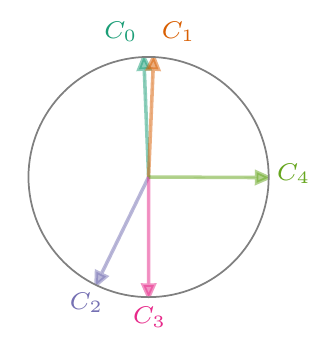}}\label{fig:embedding-example-3}}\\
  \subfloat[Sentiments]{\makebox[\textwidth][c]{\includegraphics[width=\textwidth]{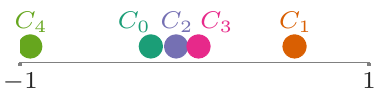}}\label{fig:embedding-example-3-sent}}
\end{minipage}%
\begin{minipage}[c]{0.75\linewidth}

\framecomment{0}{You can forget about \$16 for a while.}
\framecomment{1}{Bye-Bye twitter sweet 16.}
\framecomment{2}{\omitted{} when world leaders speak, they turn to Twitter first. \omitted{} How is it trading at \$16? \omitted{} How come Dorsey\footnote{Jack Dorsey, CEO of Twitter} can't monetize this instantaneous platform?}
\framecomment{3}{\omitted{} It's about time [for] more positive news to get it \omitted{} up again. \omitted{} Seems to have support \$16ish. \omitted{}}
\framecomment{4}{[Wall Street/CNBC] only want to pump [selective] stocks \omitted{} Twitter of China, Weibo, is selling for \$88.00 a share \omitted{}}


\end{minipage}
\caption{A subset of comments and estimated opinions for an online discussion about finance (price of Twitter stock).
  There are two distinct issues being discussed: (i) the objective price of the stock (\comment{0}, \comment{1}, \comment{2}, \comment{3}) and (ii) a subjective discussion about reasons for the supressed price (\comment{2}, \comment{4}).
\comment{0} and \comment{1} say that the price will stay below \$16 (using some humor), while \comment{2} and \comment{3} suggest that the price \emph{may} rise up.
\comment{4} suggests Wall Street/media bias against the stock and is neutral about the price of stock, while \comment{2} questions the management of the company instead.
} \label{fig:embeddings-3}
%
\end{figure*}

Finally, we take a close look into the comments, inferred multidimensional opinions and unidimensional sentiment\footnote{Comment sentiments were calculated using Convolutional Neural Networks trained on Stanford Sentiment Treebank~\cite{kim2014convolutional,socher2013recursive}.} of a discussion about politics,
shown in Figure~\ref{fig:embeddings-2}, and a discussion about finance, shown in Figure~\ref{fig:embeddings-3}.
The discussion about politics shows that, even if the lexical overlap between comments which express a similar opinion is low, \eg, \comment{0} and \comment{2} or \comment{4} and \comment{5}, 
our opinion estimation method is able to identify they are similar, as a human would do, by leveraging the judgments of the voters.
Note that, due to their low lexical overlap, it would be difficult to identify such similarity using methods based on textual analysis, as revealed by the unidimensional sentiments.
The discussion about the price of Twitter stock (see Figure~\ref{fig:embeddings-3}) shows that our method is able to capture objective opinions about the price (whether it stays at \$16 or goes up, \comment{3}, or stays down, \comment{0} and \comment{1}), along one axis and subjective opinions questioning the reason behind the price drop along a different axis (suggesting management is the reason, \comment{2}, or media bias/corruption in Wall Street, \comment{4}).
Note that \comment{2} suggests both, that the price should go up, and that the reason for the decrease is the management.
Also in this case, an analysis of the unidimensional sentiments would not reveal these rich relationships among the comments.

\vspace{-2mm}

\vspace{-2mm}
\section{Conclusion}
\label{sec:conclusion}
In this work, we have proposed a modeling framework to generate latent representations of opinions using human judgments, as measured by online voting.
As a consequence, such representations exhibit a remarkable semantic property: if two opinions are close in the latent space of opinions, it is because the
voters---the crowd---think that they are similar.
%
%
Our modeling framework is theoretically grounded and establishes an unexplored, surprising connection between opinion and voting models and the sign-rank of a matrix.
Moreover, it also provides a set of practical algorithms to both estimate the dimension of the latent space of opinions and infer where opinions expressed in comments
and held by voters lie in this space.
Experiments on a large dataset from Yahoo! News show that many discussions are multisided and avoid falling prey to demagoguery, provide insights into human judgments
and opinions, and show that our framework is able to circumvent language nuances, \eg, sarcasm and humor, by relying on human judgments.
%

%
%
%
%
%

Our work also opens up many interesting venues for future work.
For example, our measure of complexity---the dimension of the latent space of opinions---may be a good starting point to develop theoretically grounded
measures of polarization~\cite{choi2010identifying, mejova2014controversy, conover2011political} and controversy~\cite{guerra2013measure,garimella2018quantifying},
which have been lacking in the literature.
Moreover, it would be very interesting to augment our modeling framework to also incorporate, in addition to the voting data, the textual information in the comments, the identity of commenters, and their trust-worthiness.
Besides increasing the accuracy of our method, these may aid the interpretability of the dimensions as well.
Our algorithm for determining the minimum dimension under which the opinion space is able to explain the voting data exhibits weak theoretical guarantees though it
performs well on real-data.
It would be interesting to develop exact algorithm by adapting recent advances in exact sign-rank estimation~\cite{bhangale2015complexity,basri2009visibility}.
%

%
%
%
%
%

\xhdr{Acknowledgements}
We thank Mounia Lalmas, Dmitry Chistikov and Rupak Majumdar for useful discussions.

\vspace{-2mm}

\bibliographystyle{abbrv}
\bibliography{refs}

\end{document}